\documentclass[10pt]{bmc_article}    

\usepackage{cite} 
\usepackage{url}  
\usepackage{ifthen}  
\usepackage{multicol}   
\usepackage[utf8]{inputenc} 
\usepackage{enumerate}
\usepackage{amsmath,amssymb}
\usepackage{subfig}
\usepackage{graphicx}
\usepackage{epsfig}

\newboolean{publ}

\newenvironment{bmcformat}{\baselineskip20pt\sloppy\setboolean{publ}{false}}{\baselineskip20pt\sloppy}

\newcommand{\ext}{Extant}
\newcommand{\los}{GeneLoss}
\newcommand{\nul}{NoEvent}
\newcommand{\dup}{GeneDuplication}
\newcommand{\spe}{Speciation}
\newcommand{\spo}{SpeciationOut}
\newcommand{\out}{SpeciationOutside}
\newcommand{\tra}{Transfer}

\newcommand{\nd}{\noindent}
\newcommand{\ie}{\emph{i.e.}}
\newcommand{\eand}{\emph{and}}

\newcommand{\me}{E^*}
\newcommand{\de}{\mbox{\textbf{D1}}}
\newcommand{\dz}{\mbox{\textbf{D0}}}
\newcommand{\dzz}{\mbox{\textbf{D00}}}
\newcommand{\dt}{\mbox{\textbf{D2}}}
\newcommand{\dtw}{\mbox{\textbf{D12}}}
\newcommand{\se}{\mbox{\textbf{S1}}}
\newcommand{\sz}{\mbox{\textbf{S0}}}

\newtheorem{theorem}{Theorem}
\newtheorem{case}[theorem]{Case}

\begin{document}
\begin{bmcformat}


\title{Lateral Gene Transfer, Rearrangement and Reconciliation}
 

\author{
  Murray Patterson\correspondingauthor$^{1,2,3}$
  \email{Murray Patterson\correspondingauthor - Murray.Patterson@cwi.nl}
  and
  Gergely J Sz\"oll\H{o}si$^{2,4}$
  \email{Gergely Sz\"{o}ll\H{o}si - Gergely.Szollosi@univ-lyon1.fr}
  and 
  Vincent Daubin$^2$
  \email{Vincent Daubin - Vincent.Daubin@univ-lyon1.fr}
  and
  Eric Tannier\correspondingauthor$^{1,2}$
  \email{Eric Tannier\correspondingauthor - Eric.Tannier@inria.fr}
}


\address{
  \iid(1)INRIA Rh\^one-Alpes, 655 avenue de l'Europe, F-38344 Montbonnot, France\\
  \iid(2)Laboratoire de Biom\'etrie et Biologie \'Evolutive, CNRS and Universit\'e de Lyon 1, 43 boulevard du 11 novembre 1918, F-69622 Villeurbanne, France\\
  \iid(3) Centrum Wiskunde \& Informatica, Science Park 123, 1098 XG, Amsterdam, The Netherlands\\
  \iid(4) ELTE-MTA ``Lend\"ulet'' Biophysics Research Group 1117 Bp., P\'azm\'any P. stny. 1A., Budapest, Hungary
}

\maketitle


\begin{abstract}
\paragraph{Background.} 
Models of ancestral gene order reconstruction have progressively
integrated different evolutionary patterns and processes such as
unequal gene content, gene duplications, and implicitly sequence
evolution via reconciled gene trees. In unicellular organisms, these
models have so far ignored lateral gene transfer, even though it can
have an important confounding effect on such models, as well as a rich
source of information on the function of genes through the detection
of transfers of entire clusters of genes.
\paragraph{Result.} 
We report an algorithm together with its implementation, DeCoLT, that
reconstructs ancestral genome organization based on reconciled gene
trees which summarize information on sequence evolution, gene
origination, duplication, loss, and lateral transfer.  DeCoLT finds in
polynomial time the minimum number of rearrangements, computed as the
number of gains and breakages of adjacencies between pairs of genes.
We apply DeCoLT to 1099 gene families from 36 cyanobacteria genomes.
\paragraph{Conclusion.} 
DeCoLT is able to reconstruct adjacencies in 35 ancestral bacterial
genomes with a thousand genes families in a few hours, and detects
clusters of co-transferred genes.  As there is no constraint on genome
organization, adjacencies can be generalized to any relationship
between genes to reconstruct ancestral interactions, functions or
complexes with the same framework.
\paragraph{Availability.} 
\url{http://pbil.univ-lyon1.fr/software/DeCoLT/}
\end{abstract}

\ifthenelse{\boolean{publ}}{\begin{multicols}{2}}{}



\section*{Introduction}

The evolution of genomes can be explored at two different scales.  At
the chromosome level, rearrangements have been studied from the
1930's~\cite{Sturtevant1936,Sturtevant1937}, and have progressively
incorporated the possibility of unequal gene content, gene
duplications and gene losses \cite{Fertin2009,Lin2013}.  Later, but
largely independently, in the 1960's, the evolution of genomic
sequences began to be modeled~\cite{Zuckerkandl1965,Felsenstein2004},
and has more recently been extended to include the duplication, loss
and the lateral transfer of genes via the reconciliation of gene trees
with a species tree
\cite{Page1994,Maddison1998,Arvestad2003,Doyon2011,Szollosi2012}. These
two scales have only met on a few occasions through the integration of
phylogenies and rearrangements \cite{Sankoff2000,Berard2012}, using
reconciled phylogenies to account for the duplication and loss of
genes. Here, we build on these ideas and reconstruct ancestral gene
order based on reconciled phylogenies that account for gene
origination, duplication, loss, and transfer.

We propose an algorithm to simultaneously reconstruct all gene
organizations along a species phylogeny, minimizing the number of
gains and breakages of {\em adjacencies} that link consecutive genes
on chromosomes. We build upon the dynamic programming principle
proposed by B\'erard {\em et al.}~\cite{Berard2012} and extend it to
consider reconciliations (containing lateral gene transfer) produced
by Sz{\"o}ll\H{o}si {\em et al.}~\cite{Szollosi2013} as input.

We implement our algorithm naming the resulting software {\em DeCoLT},
in reference to DeCo~\cite{Berard2012} (Detection of Coevolution) with
Lateral Transfers. We examine two datasets of gene trees from a single
set of cyanobacteria species. The first set of gene trees is computed
from sequence alignements only~\cite{Guindon2010}, and the second one
is computed by a species tree aware method~\cite{Szollosi2013a}.

Our method and the efficiency of the computation is based on the
hypothesis that adjacencies evolve independently from each
other. While extant genomes consist either of a single or a relatively
small number of linear or circular chromosomes, this hypothesis
implies that reconstructed ancestral genomes may in theory exhibit
more complex arrangements. For example an ancestral gene may be
involved in more then two adjacencies, or a large number may have only
a single adjacent gene. In the cyanobacteria dataset, extant genomes
are all circular, and the ancestral genomes inferred by DeCoLT are
also close to being circular with only a few deviations. Most
deviations result from the absence of signal to reconstruct genomes in
deep ancestors, but some are caused by errors in gene trees, leading
to errors in ancestral gene contents. We observe that ancestral genome
organizations computed from gene trees that are based on both the
species tree and the sequence are closer to being circular than those
computed from gene trees based on sequence alone. This validates the
reconstruction principle we present here and confirms that species
tree aware methods produce more accurate gene trees.

\section*{Context}

A {\em dated species tree} is a rooted binary tree whose leaves are
the {\em extant genomes} and internal nodes, the {\em ancestral
  genomes}, are totally ordered. The time interval between two
consecutive internal nodes of a species tree defines a {\em time
  slice} (see Figure \ref{fig:method} for an example, where branches
leading to $A$, $B$ or the ancestor of $C$ and $D$ overlap two time
slices while the others overlap one).  Genomes contain a set of {\em
  genes} and a set of {\em adjacencies}, which are pairs of genes, the
genes being the two {\em extremities} of the adjacency.  In extant
genomes, an adjacency means that two genes are immediately
consecutive, with no other gene between them on the chromosome
(regardless of their physical distance) so every gene belongs to
exactly two adjacencies.

Genes of all genomes are partitioned into {\em homologous families},
and each family is organized in a {\em gene tree}, which is a rooted
tree whose nodes are the genes, describing the pattern of descent
within a family. Gene trees are {\em reconciled} with the species
tree, which means that nodes and branches of gene trees are annotated
to account for the particular history of the gene family. Possible
events are origination (of the gene in the species tree), speciation
(genes follow the species diversification), duplication, transfer, or
loss. Transfers are the acquisition of a gene by a genome in the
species tree from a genome outside the species tree.  Indeed genes at
the origin of transfers almost always belong to unsampled or extinct
species~\cite{Szollosi2013}. That is why speciation does not only
happen at the nodes of the species tree, but a gene can also leave the
species tree by speciation, and be transferred back later (see Figure
\ref{fig:method} for an example).  Every event on a gene tree is
associated to a branch and a time slice of the species tree.

\begin{figure}
  \begin{center}
  \includegraphics[width=12cm]{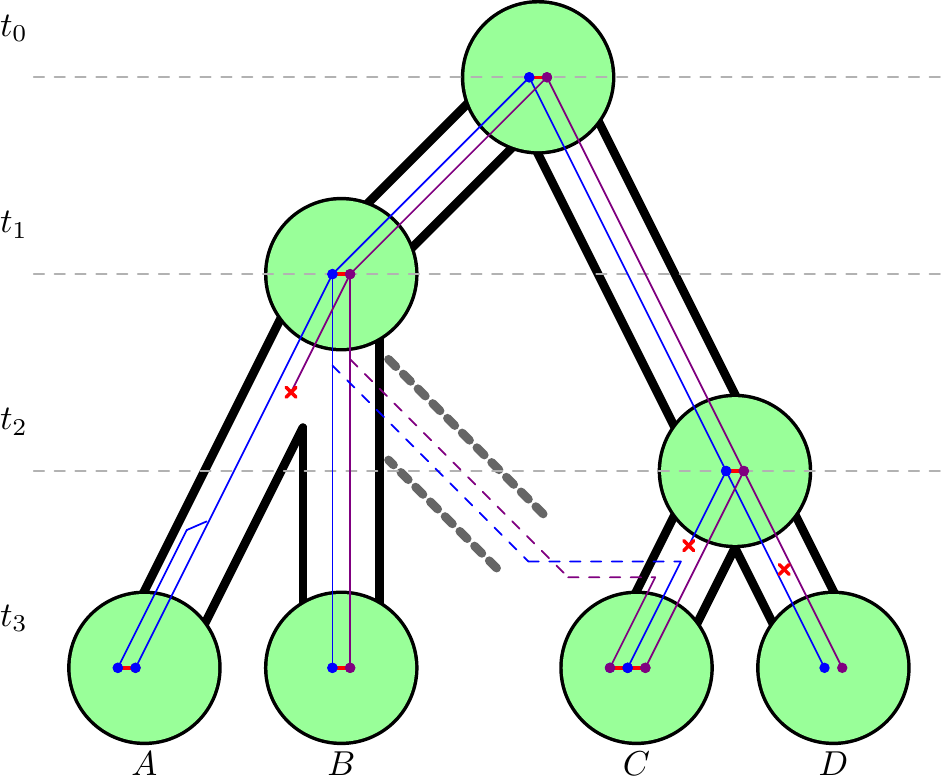}
  \caption{The evolution of an adjacency within a dated species tree,
    along reconciled gene phylogenies.  The gene trees are blue and
    purple, while red horizontal edges are adjacencies.  The time
    slices $t_0,\dots,t_3$ indicate in which order the speciation
    nodes (big green nodes) occur, and are used to localize genes in
    the species tree (a branch and a time slice give the coordinates
    of a gene or an event).  Red crosses mean gene losses, for example
    in the branches leading to $A$ or $C$ (adjacencies are lost when
    one extremity is lost), or an adjacency breakage, for example in
    the branch leading to $D$ (gene loss and adjacency breakages are
    different events, since a gene loss is not a rearrangement while a
    breakage is, and only rearrangements are counted in the objective
    function).  Here one adjacency is gained in the branch leading to
    species $C$, one is broken in the branch leading to species $D$,
    and one is transferred from the branch leading to $B$ to the
    branch leading to $C$.  The transfer implies first a speciation
    outside the species phylogeny, and then a transfer which can be in
    another time slice.  A tandem duplication in the branch leading to
    $A$ gives a new adjacency between the two copies.}
  \label{fig:method}
  \end{center}
\end{figure}

The input to our method is a dated species tree, a set of reconciled
gene trees and the set of extant adjacencies. Reconciled gene trees
yield ancestral genes (dots inside green circles in Figure
\ref{fig:method}).  The problem will be to construct the ancestral
adjacencies, given this input. In practice the input is provided by
methods and software described in Sz{\"o}ll\H{o}si {\em et al.}'s
trilogy~\cite{Szollosi2012,Szollosi2013,Szollosi2013a}.  The first
paper of this trilogy explains how to find the dated species tree, the
second one how to reconcile gene trees taking extinct or unsampled
species into account, and the third one how to reconstruct species
aware gene trees.

We construct ancestral adjacencies in a manner that minimizes the
number of rearrangements along the species phylogeny. This is computed
as the smallest number of gains and breakages of adjacencies necessary
to explain all extant adjacencies. For example, if we infer no
ancestral adjacencies at all, then the value of this objective
function is proportional to the number of extant adjacencies, because
all of them are gained independently.  If we propose an adjacency in
an ancestral genome which is a common ancestor of a set of
adjacencies, then the value decreases because all adjacencies in that
set are explained by a unique gain (see Figure \ref{fig:method}, where
three extant adjacencies can be explained with two gains and a
breakage).

We then have to describe how an ancestral adjacency {\em propagates}
within reconciled gene trees so that it can be recognized as an
ancestor of an extant one.

\section*{Propagation rules}

There is an adjacency between two genes only if they are in the same
species (extant or ancestral) at the same time slice.  If there is an
adjacency between two ancestral genes $a$ and $b$, it is propagated to
the descendants, in the absence of rearrangements, following the rules
described in Table \ref{tab:propagation}, according to the events
happening to genes $a$ and $b$.

\begin{center}
  \begin{tabular}[c]{|c|c|c|c|c|}
    \hline
    \textbf{event for $a$} & \textbf{event for $b$} & \textbf{adjacencies in} & \textbf{graphical} & \textbf{recurrence}\\
    && \textbf{descendants} & \textbf{depiction} & \textbf{rules}\\
    \hline
    \hline
    $\los$ & any event & $\emptyset$&\includegraphics[height=1.5cm]{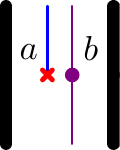}&\ref{c:los-any}\\
    \hline
    no event & no event &$a_1b_1$&\includegraphics[height=2cm]{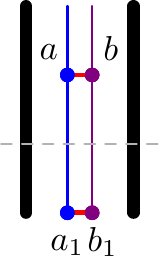}&\ref{c:nul-nul}\\
    \hline
    no event & $\dup$ &$ab_1$ or $ab_2$&\includegraphics[height=2cm]{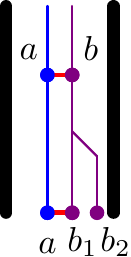}&\ref{c:nul-dup}\\
    \hline
    $\dup$ & $\dup$ &$a_1b_1$ and $a_2b_2$ or&\includegraphics[height=2cm]{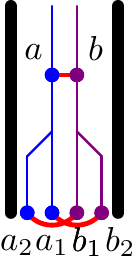}&\ref{c:dup-dup}\\
    && $a_1b_2$ and $a_2b_1$&&\\
    \hline
    $\spe$ & $\spe$ &$a_1b_1$ and $a_2b_2$&\includegraphics[height=2.5cm]{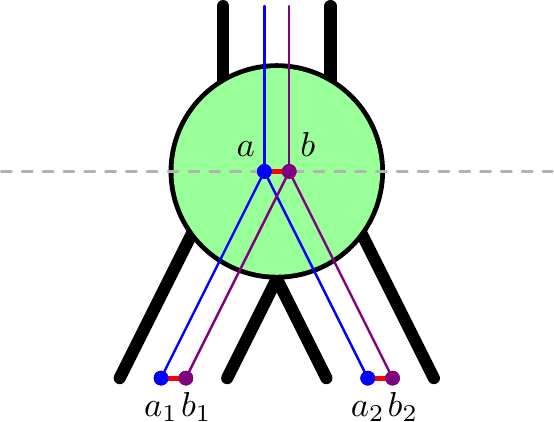}&\ref{c:spe-spe}\\
    \hline
  \end{tabular}
\end{center}
\hfill see next page\dots\\

\begin{table}
  \begin{center} 
    \begin{tabular}[c]{|c|c|c|c|c|}
      \hline
      \textbf{event for $a$} & \textbf{event for $b$} & \textbf{adjacencies in} & \textbf{graphical} & \textbf{recurrence}\\
      && \textbf{descendants} & \textbf{depiction} & \textbf{rules}\\
      \hline
      \hline
      no event & $\spo$ &$ab_1$&\includegraphics[height=2.5cm]{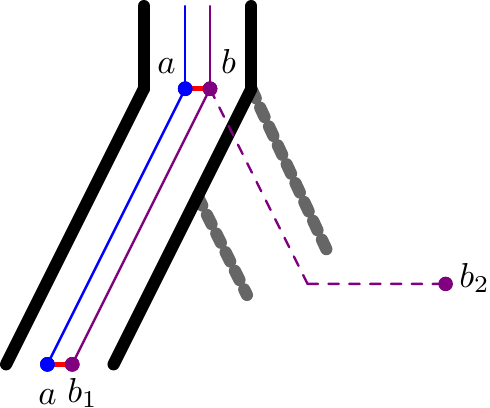}&\ref{c:nul-spo}\\
      \hline
      $\spo$ & $\spo$ &$a_1b_1$ and $a_2b_2$ or&\includegraphics[height=2.5cm]{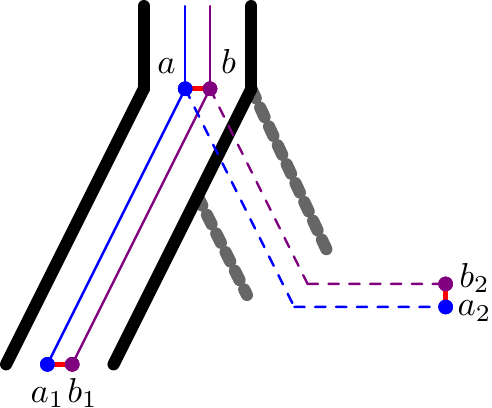}&\ref{c:spo-spo}\\
      && $a_1b_2$ and $a_2b_1$&&\\
      \hline
      no event & $\out$ & $ab_1$ or $ab_2$&\includegraphics[height=2.5cm]{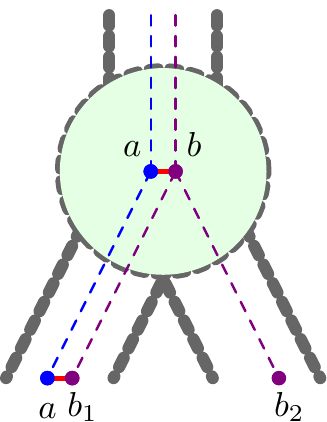}&\ref{c:out-out}\\
      \hline
      $\out$ & $\out$ &$a_1b_1$ and $a_2b_2$&\includegraphics[height=2.5cm]{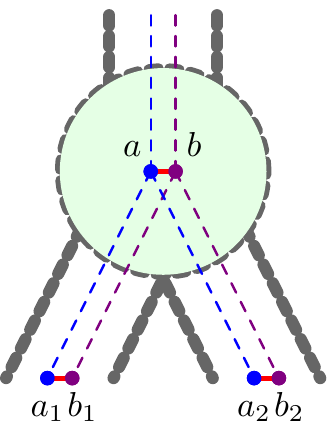}&\ref{c:out-out}\\
      \hline
      $\tra$ & any event &$ab$&\includegraphics[height=2cm]{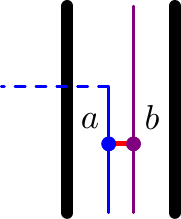}&\ref{c:tra-any}\\
      \hline
      $\tra$ & $\tra$ &$ab$&\includegraphics[height=2cm]{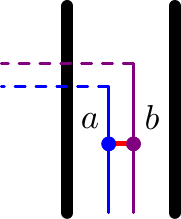}&\ref{c:tra-any}\\
      \hline
    \end{tabular}
  \end{center}
  \caption{Propagation rules for the adjacencies: a function of the events happening
    to their extremities.  We denote $a_1$ and $a_2$ (resp., $b_1$ and $b_2$) as the children of $a$ (resp., $b$).  Recurrence rules then follow
    exactly the propagation rules, adding the possibility of
    rearrangements at each step.  $\spe$ means speciation inside the
    species tree, $\spo$ means the parent is inside and one child is
    inside and the other child is outside the species tree, while $\out$
    means that the parent (and possibly the children) are outside.
    Note that ``no event'' is any non-leaf event (leaves of gene trees are labeled with $\ext$, see the recurrence rules), while ``any event'' is any non-leaf or leaf event.
    Note that ``no event'' cannot occur with speciation, since both trees are reconciled (see Figure~\ref{fig:method}), while ``no event''  happens with ``no event'' when both genes change time slice without any event (for example, the change from $t_1$ to $t_2$ in the branch between the root and its right child in Figure~\ref{fig:method}).}
  \label{tab:propagation}
\end{table}

A {\em history} is a set of ancestral and extant adjacencies. In a
history, any adjacency which does not have a parent identified by the
propagation rules yields an adjacency {\em gain}.  A {\em breakage} is
inferred when an adjacency is present in the history but one of its
descendants according to the propagation rules is not.  The {\em cost}
of a history is the number of gains and breakages it yields. We are
computing a minimum cost history.

\section*{Algorithm}

We compute a minimum cost history by writing a dynamic programming
algorithm following the propagation rules and adding adjacency gains
and breakages with costs that are considered in the optimization.  In
order to solve a more general problem and to present the recurrence
formulas more clearly, gains and breakages are assigned a cost, which
could be different, and we minimize on the number of events weighted
by their cost. In practice we always use the algorithm with equal
costs, thus minimizing the sum of the number of gains and breakages.

\subsection*{Classes of adjacencies}

Two adjacencies are {\em homologous} with respect to a particular
history if they descend from a common ancestor following the
propagation rules. Homology of adjacencies is an equivalence
relation. We first state a necessary condition for a set of
adjacencies to be homologous in order to restrict the search space for
homology.  

Two extant adjacencies $a_1b_1$ and $a_2b_2$ are {\em possibly
  homologous} if there are two ancestral genes $a$ and $b$ of an
ancestral genome $G$, such that $a$ (resp., $b$) is an ancestor of
$a_1$ and $a_2$ (resp., $b_1$ and $b_2$).  This simply tells us that
in order to find a common ancestor of two adjacencies, there has to
exist two genes being the extremities of this adjacency. So if two
adjacencies are homologous with respect to a particular history then
they are possibly homologous (the definition of possible homology is
independent from any history).  Possible homology, defined on two
adjacencies, is obviously a symmetric and reflexive relation.  It is
also transitive, partitioning the set of extant adjacencies into
equivalence classes.

Consequently, homology can be searched within a class. For each class
$\{a_1b_1,\dots,a_kb_k\}$, there are two genes $a$ and $b$, such that
$a$ (resp. $b$) is an ancestor of all $a_i$ (resp., $b_i$). Among the
possible such genes $a$ and $b$ for a class, we call the highest
distinct ones the {\em roots} of that class. We then work with the
disjoint subtrees rooted by $a$ and $b$, and find a history following
the propagation rules for all adjacencies whose extremities are
descendants of $a$ and $b$.  Hence, it is sufficient to search within
pairs of trees to construct a history.

\subsection*{Recurrence formulas within one class}

For any two gene tree nodes $a$ and $b$, for which $s(a)=s(b)$, let
$c_1(a,b)$ be the minimum cost of a history for the two gene subtrees
rooted at $a$ and $b$, assuming there is an adjacency between $a$ and
$b$, and let $c_0(a,b)$ be the minimum cost of a history for two gene
subtrees rooted at $a$ and $b$, assuming there is no adjacency between
$a$ and $b$. The values of $c_1(a,b)$ and $c_0(a,b)$ are recursively
computed, according to the events annotating $a$ and $b$. Thus we have
to enumerate all cases.

Given a node $u$, $u_1$ is the first (or only child of $u$ in the case
that $u$ has only one child), while $u_2$ is the second child. We
write $E(u)$ to denote the event at node $u$, where $E(u)=\ext$ when
$u$ is a leaf of a gene tree corresponding to an extant gene.

\begin{case} \label{c:ext-ext}
  $E(a)=\ext$~\eand~$E(b)=\ext$ (both nodes are leaves).
\end{case}
In this case, if $ab$ is an adjacency then $c_1(a,b) = 0$ and
$c_0(a,b) = \infty$, else $c_1(a,b) = \infty$ and $c_0(a,b) = 0$.\\

\begin{case} \label{c:los-any}
  $E(a)=\los$ (one of the genes is lost, any event may happen to the other).
\end{case}
In this case $c_1(a,b) = 0$ and $c_0(a,b) = 0$.\\

\begin{case} \label{c:nul-nul}
  $E(a)=\nul$~\eand~$E(b)=\nul$ (both gene trees are changing time slice without any event).
\end{case}
In this case $c_1(a,b) = \min\{ c_1(a_1,b_1), c_0(a_1,b_1)+
C(Break)\}$ and $c_0(a,b) = \min\{c_0(a_1,b_1), c_1(a_1,b_1)+
C(Gain)\}$.

\begin{case} \label{c:nul-dup}
  $E(a)\in\{\ext,~\nul,~\spe,~\spo\}$~\eand~$E(b)=\dup$
\end{case}
In this case we suppose that the duplication of $b$ happens before any
event in the gene tree containing $a$.  Here, $c_1(a,b) = \de$, and
$c_0(a,b) = \dz$, where

$$
\de = \min \left\{
  \begin{array}{l}
    c_1(a,b_1) + c_0(a,b_2),\\
    c_0(a,b_1) + c_1(a,b_2),\\
    c_1(a,b_1) + c_1(a,b_2) + C(Gain),\\
    c_0(a,b_1) + c_0(a,b_2) + C(Break)\\
  \end{array}
\right.
\hspace*{.5cm}
\dz = \min \left\{
  \begin{array}{l}
    c_0(a,b_1) + c_0(a,b_2),\\
    c_0(a,b_1) + c_1(a,b_2) + C(Gain),\\
    c_1(a,b_1) + c_0(a,b_2) + C(Gain),\\
    c_1(a,b_1) + c_1(a,b_2) + 2*C(Gain)\\
  \end{array}
\right.
$$

\begin{case} \label{c:dup-dup}
  $E(a)=\dup$~\eand~$E(b)=\dup$.
\end{case}
In this case $c_1(a,b) = \min(\de,~\dtw,~\dtw)$ where

\nd $\de$ (defined in Case~\ref{c:nul-dup}) is the cost in the case
where the $a$ duplication comes first.

\nd $\dt$ is the cost in the case where the $a$ duplication comes
first:

$$
\dt = \min \left\{
  \begin{array}{l}
    c_1(a_1,b) + c_0(a_2,b),\\
    c_0(a_1,b) + c_1(a_2,b),\\
    c_1(a_1,b) + c_1(a_2,b) + C(Gain),\\
    c_0(a_1,b) + c_0(a_2,b) + C(Break)\\
  \end{array}
\right.
$$

\nd $\dtw$ is the cost in the case where the $a$ and $b$ duplications
are simultaneous: $\dtw = \min$ (over all 16 of the following cases):

$$
\left\{
  \begin{array}{l}
    (1)~~c_1(a_1,b_1) + c_1(a_2,b_2) + c_0(a_1,b_2) + c_0(a_2,b_1),\\
    (2)~~c_1(a_1,b_1) + c_1(a_2,b_2) + c_0(a_1,b_2) + c_1(a_2,b_1) + C(Gain),\\
    (3)~~c_1(a_1,b_1) + c_1(a_2,b_2) + c_1(a_1,b_2) + c_0(a_2,b_1) + C(Gain),\\
    (4)~~c_1(a_1,b_1) + c_1(a_2,b_2) + c_1(a_1,b_2) + c_1(a_2,b_1) + 2*C(Gain),\\
    (5)~~c_1(a_1,b_1) + c_0(a_2,b_2) + c_0(a_1,b_2) + c_0(a_2,b_1) + C(Break),\\
    (6)~~c_1(a_1,b_1) + c_0(a_2,b_2) + c_0(a_1,b_2) + c_1(a_2,b_1) + C(Gain) + C(Break),\\
    (7)~~c_1(a_1,b_1) + c_0(a_2,b_2) + c_1(a_1,b_2) + c_0(a_2,b_1) + C(Gain) + C(Break),\\
    (8)~~c_0(a_1,b_1) + c_1(a_2,b_2) + c_0(a_1,b_2) + c_0(a_2,b_1) + C(Break),\\
    (9)~~c_0(a_1,b_1) + c_1(a_2,b_2) + c_0(a_1,b_2) + c_1(a_2,b_1) + C(Gain) + C(Break),\\
    (10)~~c_0(a_1,b_1) + c_1(a_2,b_2) + c_1(a_1,b_2) + c_0(a_2,b_1) + C(Gain) + C(Break),\\
    (11)~~c_0(a_1,b_1) + c_0(a_2,b_2) + c_1(a_1,b_2) + c_1(a_2,b_1),\\
    (12)~~c_0(a_1,b_1) + c_1(a_2,b_2) + c_1(a_1,b_2) + c_1(a_2,b_1) + C(Gain),\\
    (13)~~c_1(a_1,b_1) + c_0(a_2,b_2) + c_1(a_1,b_2) + c_1(a_2,b_1) + C(Gain),\\
    (14)~~c_0(a_1,b_1) + c_0(a_2,b_2) + c_1(a_1,b_2) + c_0(a_2,b_1) + C(Break),\\
    (15)~~c_0(a_1,b_1) + c_0(a_2,b_2) + c_0(a_1,b_2) + c_1(a_2,b_1) + C(Break),\\
    (16)~~c_0(a_1,b_1) + c_0(a_2,b_2) + c_0(a_1,b_2) + c_0(a_2,b_1) + 2*C(Break)\\
  \end{array}
\right.
$$

\nd And, $c_0(a,b) = \dzz$, where

$$
\dzz = \min \left\{
  \begin{array}{l}
    \dz\\
    c_0(a_1,b) + c_0(a_2,b),\\
    c_0(a_1,b) + c_1(a_2,b) + C(Gain),\\
    c_0(a_1,b) + c_0(a_2,b) + C(Gain),\\
    c_1(a_1,b) + c_1(a_2,b) + 2*C(Gain)\\
  \end{array}
\right.
$$

\begin{case} \label{c:spe-spe}
  $E(a)=\spe$~\eand~$E(b)=\spe$.
\end{case}
We assume without loss of generality that $s(a_1) = s(b_1)$ and
$s(a_2) = s(b_2)$.  Here, $c_1(a,b) = \se$ and $c_0(a,b) = \sz$, where

$$
\se = \min \left\{
  \begin{array}{l}
    c_1(a_1,b_1) + c_1(a_2,b_2),\\
    c_1(a_1,b_1) + c_0(a_2,b_2) + C(Break),\\
    c_0(a_1,b_1) + c_1(a_2,b_2) + C(Break),\\
    c_0(a_1,b_1) + c_0(a_2,b_2) + 2*C(Break)\\
  \end{array}
\right.
\hspace*{.5cm}
\sz = \min \left\{
  \begin{array}{l}
    c_0(a_1,b_1) + c_0(a_2,b_2),\\
    c_1(a_1,b_1) + c_0(a_2,b_2) + C(Gain),\\
    c_0(a_1,b_1) + c_1(a_2,b_2) + C(Gain),\\
    c_1(a_1,b_1) + c_1(a_2,b_2) + 2*C(Gain)\\
  \end{array}
\right.
$$

\begin{case} \label{c:nul-spo}
  $E(a)\in\{\ext,~\nul,~\spe\}$~\eand~$E(b)=\spo$.
\end{case}
In this case $c_1(a,b) = c_1(a,b_1)$ and $c_0(a,b) =
c_0(a,b_1)$.\\

\begin{case} \label{c:spo-spo}
  $E(a)=\spo$~\eand~$E(b)=\spo$.
\end{case}
We assume without loss of generality that $a_1$ (resp., $b_1$) is the
child that remains inside the species tree, while $a_2$ (resp., $b_2$)
is the child that leaves the tree.  In this case, $c_1(a,b) =
c_1(a_1,b_1) + \min\{c_1(a_2,b_2), c_0(a_2,b_2)+C(Break)\}$ and
$c_0(a,b) = c_0(a_1,b_1) + \min\{c_0(a_2,b_2),
c_1(a_2,b_2)+C(Gain)\}$.

\begin{case} \label{c:out-out}
  $E(b)=\out$ (one of the genes is outside of the species tree, any
  event may happen to the other).
\end{case}
Here, we compute values for $a$ and $b$ only if
$\sigma(a)$\footnote{$\sigma$ (resp., $\me$) is another type of
  species (resp., event) assignment function like $s$ (resp., $E$)
  that is defined during a preprocessing step of the backtracking
  procedure, which is described in the next section.  This rule does
  not become effective until this step is executed.}, or $\sigma(b)$
are defined, and in this case compute values only if
$\sigma(a)=\sigma(b)$.  Then, if $\me(a)=\nul$ and $\me(b)=\spe$, then
$c_1(a,b)=\de$ and $\me(b)=\dz$ (both defined in
Case~\ref{c:nul-dup}).  Else, $\me(a)=\me(b)=\spe$ and then
$c_1(a,b)=\se$ and $c_0(a,b)=\sz$ (both defined in
Case~\ref{c:spe-spe}).

\begin{case} \label{c:tra-any}
  $E(a)=\tra$ (one of the genes is a transfer, any event may happen to the other).
\end{case}
In this case, since $s(a)=s(b)$, the values $c_1(a,b)$ and $c_0(a,b)$
have already been computed recursively, and hence they are these
values.  Observe that if $E(b)\not\in\{\tra,\los\}$ then $(a,b)$ will
form the root of an equivalence class.

\subsection*{Backtracking Procedure}

First, the dynamic programming matrix $M[a,b]$ containing a cell for
each pair $(a,b)$ of nodes in the respective gene trees is created by
following the recurrence rules for each equivalence class.  However,
all of the nodes $u$ such that $E(u)=\out$ are outside of the species
tree, and hence $s(u)$ is undefined.  This means that $M[a',b']$ for
such pairs $(a',b')$ where $E(a)=E(b)=\out$ have not computed in $M$,
and so the recurrence rules cannot track the propagation of a possible
adjacency in the species tree above such a pair $(a',b')$ to a pair
below, that has been co-transferred back into the species tree.  We
now present a preprocessing step that artificially assigns an a
species and an event to such nodes $(a',b')$ to allow the recurrence
rules (\ie, Case~\ref{c:out-out}) to be able to track such
propagation.  After this step, the recurrence rules can be applied
again, with Case~\ref{c:out-out} in effect, to obtain the full $M$.
Then, a classical backtracking procedure can be used to reconstruct
ancestral adjacencies along a minimum path in $M$.  This preprocessing
step is done in a manner that guarantees that the tracking of such
propagation will result in a minimum cost history.  We now detail this
preprocessing step.

For each cell $M[a,b]$ where $E(a)=E(b)=\spo$, we do a (depth-first)
search to find all nodes $a'$ (resp., $b'$) below $a$ (resp., $b$)
such that $E(a')=$ (resp., $E(b')$) $\tra$.  since each $a'$ (resp.,
$b'$) is a point where the $\spo$ event at $a$ (resp., $b$) returns to
the species tree, let $M'$ be the set of all such pairs $(a',b')$
(\ie, cells in $M$) such that $s(a')=s(b')$.  It is only these
``frontier'' cells $M'$ that have the possibility to connect $(a,b)$
with the lower parts of the respective trees (the lower parts rooted
precisely at each $(a,b)\in M'$), through the propagation of an
adjacency.  Other such descendants of $(a,b)$ can never receive
adjacencies from $(a,b)$, so we treat them as losses and consider them
no further.

Now, let $T_a$ and $T_b$ be the subtrees of the trees containing $a$
(resp., $b$), rooted at $a$ (resp., $b$) with the leaves $a'$ (resp.,
$b'$), such that $(a',b')\in M'$.  It is the propagation of
adjacencies within this tree we need to analyze in order to most
parsimoniously (in terms of gains and breakages of adjacencies)
connect $(a,b)$ with its frontier cells $M'$.  Since all internal
nodes $u\in T_a$ (resp., $v\in T_b$), \ie, nodes of degree larger than
one have $E(u)=$ (resp., $E(v)=$) $\out$, it follows that $s(u)$ and
$s(v)$ are undefined, and so there is no restriction on possible
adjacencies between any such pair $(u,v)$ in the propagation from
$(a,b)$ down to the frontier $M'$.  In constructing a minumum cost
history that explains all extant adjacencies, we may hence assume the
most parsimonious of such scenarios.  One way to do this is to impose
an artificial species subtree topology $T_S$ that both $T_a$ and $T_b$
must follow in this propagation, in such a way that recurrence rules
(\ie, Case~\ref{c:out-out}) will allow the computation of a minimum
cost history.  We hence take this approach in the following.

If the topology of $T_a$ was identical to $T_b$, it is easy to see
that we could choose $T_S$ to be $T_a$.  The topology of $T_b$ is
different from $T_a$ in general, however we can still choose $T_S$ to
be $T_a$, and then we can force $T_b$ to follow (\ie, we
\emph{reconcile} $T_b$ in) $T_S$ in a manner that minimizes the number
of duplications of nodes in $T_b$.  This is because such a duplication
is the only situation that would invoke line~3 of $\de$ in
Case~\ref{c:out-out} during the computation of the history--no other
lines of Case~\ref{c:out-out} have a $Gain$.  Such a reconciliation
that achieves this is the LCA reconciliation~\cite{Goodman1979}, so we
use this.

We use this topology $T_S$ to assign an artificial species and event
to each internal node of $T_a$ and $T_b$ so that the recurrence
(Case~\ref{c:out-out}) may follow this pair of trees.  Since the
topology of $T_a$ is that of $T_S$, we simply set $\sigma(u)$ to some
unique integer, and $\me(u)=\spe$ for each $u\in T_a$.  We then set
$\sigma(v)$ for $v\in T_b$ to $\sigma(u)$ for the $u$ that $v$ maps to
in the LCA reconciliation.  The reconciliation of $T_b$ also
introduces duplication and loss events to $T_b$.  For either such
event, we set $\me(v)=\nul$ for the corresponding nodes, and then for
the remainder of the ($\spe$) nodes $v'$, we set $\me(v')=\spe$.  Now
that we have assigned these artificial species and events to the these
$\out$ nodes, we can simply apply the recurrence rules again (this
time making Case~\ref{c:out-out} effective) on the trees to compute
the full $M$.  Then, after applying the classical backtracking
procedure on $M$, the optimal (minimum cost) history is then obtained
by choosing the minimum among $c_1+C(Gain)$ and $c_0$ on the roots of
all classes.

\subsection*{Complexity}

Let $m$ be the number of gene trees, and $n$ be the maximum number of
genes in a gene tree.  Since there are as many as $n$ time slices
$t_0,\dots,t_n$ (see Figure~\ref{fig:method}) for any tree, there can
be as many as $n^2$ events (most of them are $\nul$ events) in a given
tree.  The number of equivalence classes is $O(m^2)$, and hence there
are $O(m^2n^4)$ comparisons computed during the initial construction
of dynamic programming matrix $M$ before the preprocessing step.  The
preprocessing step, involving (linear-time computable) LCA
reconciliation~\cite{Goodman1979}, for a pair of trees takes time
$O(n)$, for an overall time of $O(m^2n)$.  The second run of the
recurrence rules after the preprocessing step is again $O(m^2n^4)$,
for an overall (polynomial) running time of $O(m^2n^4)$.

In practice, the number of equivalence classes is much smaller--closer
to $m$ than $O(m^2)$, and the majority of the $O(n^2)$ events each
tree are $\nul$ events.  On the cyanobacteria dataset of ($m=$) 1099
families from ($n=$) 36 genomes, our implementation, DeCoLT, of this
algorithm constructed the adjacencies in under 3 hours on a standard
desktop computer.

\section*{Cyanobacteria ancestral genomes}

The algorithm has been implemented and run on two datasets. They both
have the same species tree (depicted in Figure \ref{fig:phylogeny}),
on the same set of 36 extant genomes from cyanobacteria and the same
extant adjacencies.

They differ by their set of gene trees. One of them is the {\em
  sequence trees}, which are maximum likelihood trees constructed from
a model of sequence evolution using multiple alignments of protein
sequences of extant genes from each family, taken
from~\cite{Szollosi2012}.
\begin{figure}
  \begin{center}
    \includegraphics[width=12cm]{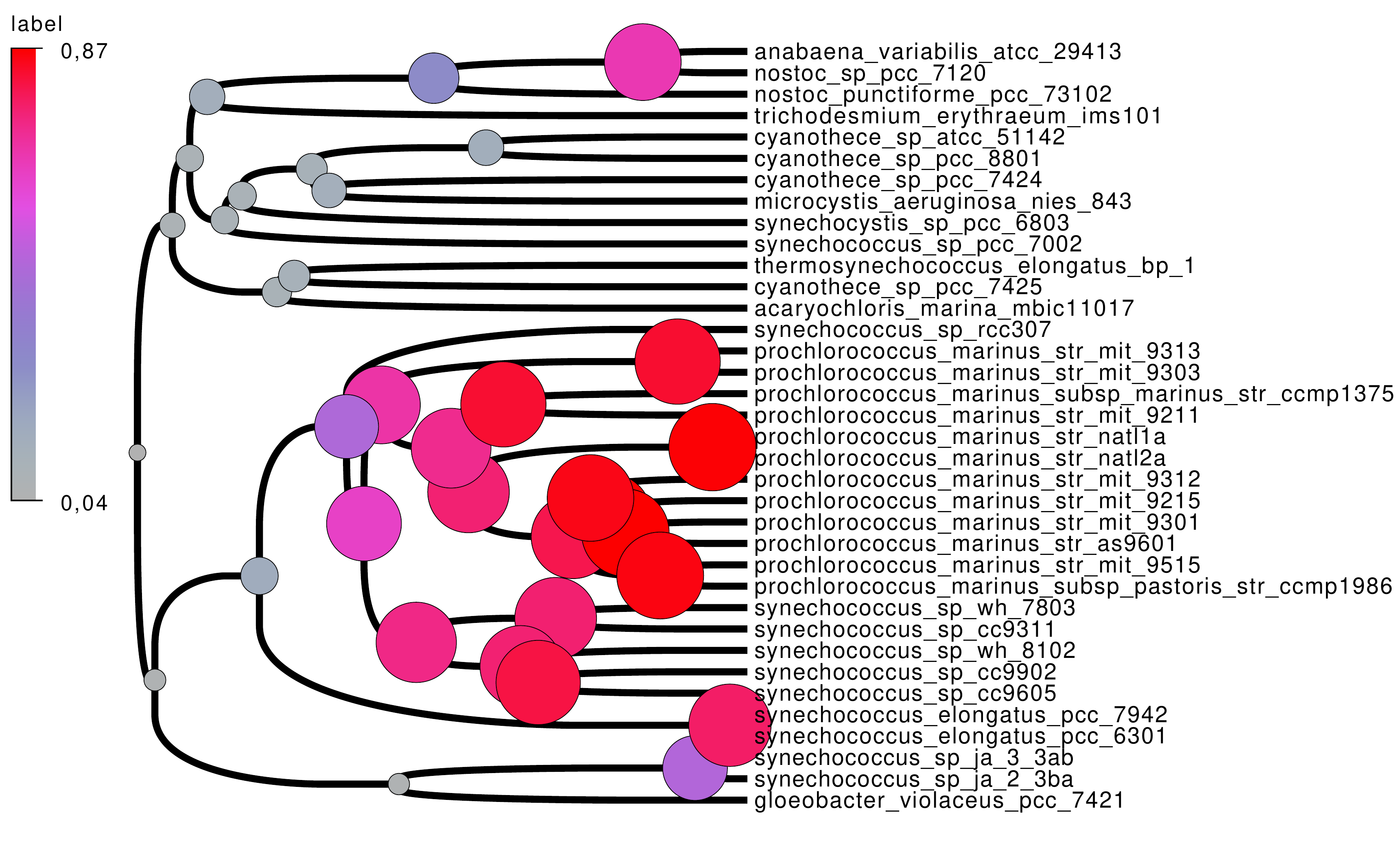}
    \caption{Cyanobacteria dated species tree. The size (area) and
      colours of internal nodes is the ratio of the number of
      adjacencies over the number of genes in every ancestral
      species.}
    \label{fig:phylogeny}
  \end{center}
\end{figure}
The other is the {\em ALE} trees, which are maximum likelihood trees
constructed from a model of sequence evolution in conjunction with a
birth and death branching model to account for origination,
duplication, transfer and loss, taken from~\cite{Szollosi2013a}.  As
transfers are very likely to involve lineages outside any given
phylogeny~\cite{Szollosi2013}, reconciled trees have nodes leaving the
species tree ($\spo$) and nodes transferring to the species tree
($\tra$).

For both datasets, ancestral adjacencies were computed using
DeCoLT. The degree of each ancestral gene (the number of adjacencies
it belongs to) was computed, and we then plotted the proportion of
ancestral genes having degree $k$ for $k$ between 0 and 6
(Figure~\ref{fig:adjacencies}).
\begin{figure}
  \begin{center}
    \includegraphics[width=8cm]{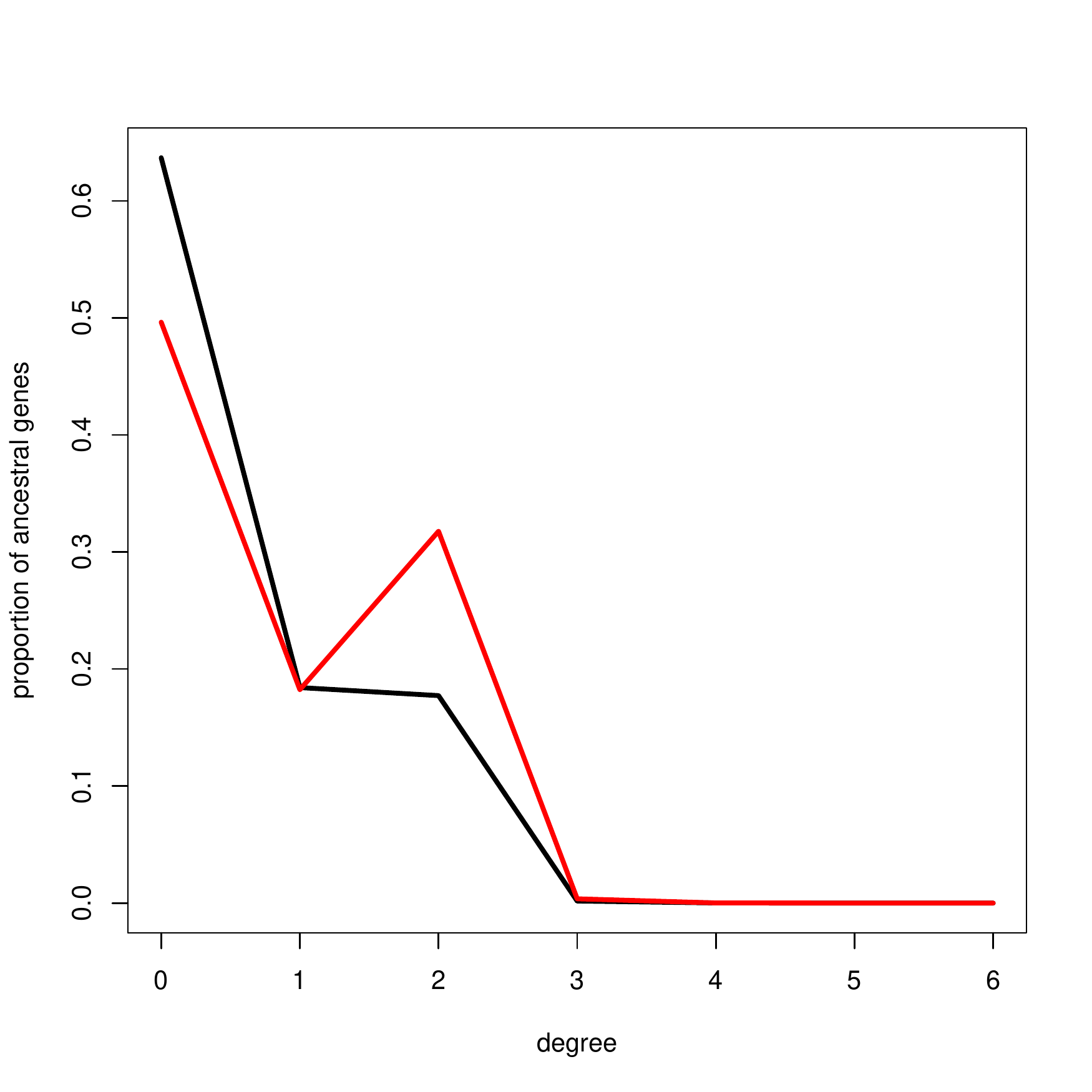}
    \caption{On the $x$ axis is the degree of a gene, that is, the
      number of adjacencies it belongs to, and on the $y$ axis there
      is the proportion of genes with this degree. In black, there are
      the values for the sequence trees and in red for ALE trees.}
    \label{fig:adjacencies}
  \end{center}
\end{figure}

There are almost no genes with degree larger than 2 in either
dataset. The proportion of genes with degree 2 increases from $17\%$
for sequence trees to $31\%$ for ALE trees.  This means that we: (i)
accurately reconstruct ancestral adjacencies because they all have a
circular structure; and (ii) the quality of gene trees nearly doubles
the resolution of ancestral genomes. Finally, having only $31\%$ of
ancestral genes with degree two means that a large part of the gene
order signal is lost in this very deep branch. However, this is not
the case for ancestral genomes. The size of the nodes on Figure
\ref{fig:phylogeny} indicates the ratio between the number of
adjacencies and the number of genes in each genome (the ideal ratio is
1).  In the {\em Prochlorococcus} clade, over 80\% of the genomes are
reconstructed whereas it drops to nearly 0\% in deeper nodes.

We found that 64 clusters of genes were co-transferred: transferred
adjacencies were detected, as well as 28 clusters of co-duplicated
genes during the evolution of cyanobacteria. Most are simply pairs of
genes, but there is a cluster of 4 co-transferred genes, four clusters
of 3 co-transferred genes, and two clusters of 3 co-duplicated genes.

\section*{Discussion}

The optimizing property of the algorithm follows from the exact
translation of the propagation rules into the recurrence formulas,
adding all possibilities of rearrangements every time. It is a
generalization of the reconstruction of discrete ancestral characters
solved by Sankoff-Rousseau type algorithms~\cite{Sankoff1975}.  To see
this, one can observe that our framework is strictly equivalent to the
Sankoff-Rousseau algorithm~\cite{Sankoff1975} in the case where there
are no events in the trees.

Further improvements in the method would consist in adding the
possibility of homolog replacement when a gene is transferred: for the
moment any transfer yields rearrangements whereas some genes might
replace an homologous one, keeping the gene order unchanged. We could
also think of avoiding rearrangements caused by origination and losses
of genes, which, for the moment, necessarily yield several adjacency
gains and losses.

Future work will also consist in deriving function information from
co-transfers, and trying the same principles on other kinds of
relations than adjacencies, starting for example from the relation
between domains forming the same gene.

\section*{Author's contributions}

MP, GS, VD and ET devised the algorithm. MP programmed the software.
ET wrote the paper.

\section*{Acknowledgements}
  \ifthenelse{\boolean{publ}}{\small}{}

Thanks to S\`everine B\'erard and Thomas Bigot.  MP is funded by a
Marie Curie Fellowship from the Alain Bensoussan program of ERCIM and
and GS is funded by the Marie Curie Fellowship 253642 "Geneforest",
and the Albert Szent-Gy\"orgyi Scholarship A1-SZGYA-FOK-13-0005.  This
work is funded by the Agence Nationale pour la Recherche, Ancestrome
project ANR-10-BINF-01-01.


\newpage
{\ifthenelse{\boolean{publ}}{\footnotesize}{\small}
 \bibliographystyle{bmc_article}  
  \bibliography{references.bib}} 


\ifthenelse{\boolean{publ}}{\end{multicols}}{}

\end{bmcformat}
\end{document}